\documentclass[]{article}

\usepackage{amsmath}
\usepackage{amsthm,amssymb}
\usepackage{color}
\usepackage{authblk}

\usepackage[numbers]{natbib}
\usepackage[T1]{fontenc}
\usepackage[utf8]{inputenc}
\usepackage{subfig}
\usepackage{graphicx}

\providecommand{\keywords}[1]
{
  \small	
  \textbf{\textit{Keywords---}} #1
}

\title{Benford's laws tests on S\&P500 daily closing values and the corresponding daily  log-returns \\both point to huge non-conformity }

\author[1,2,3]{Marcel Ausloos}
\author[4]{Valerio Ficcadenti\thanks{Corresponding author email: ficcadv2@lsbu.ac.uk}}
\author[4]{Gurjeet Dhesi}
\author[5]{Muhammad Shakeel}
\affil[1]{School of Business,  College of Social Sciences, Arts, and Humanities, University of  Leicester, Brookfield, Leicester,   LE2 1RQ, United Kingdom}  
\affil[2]{Department of Statistics and Econometrics, Bucharest University of Economic Studies, Calea Dorobantilor 15-17, 010552 Sector 1, Bucharest, Romania}
\affil[3]{Group of Researchers for Applications of Physics in Economy and Sociology (GRAPES), Rue de la belle jardini\`ere, 483,  B-4031 Angleur, Li\`ege, Belgium} 
\affil[4]{ School of Business,  London South Bank University, 103 Borough Road, SE1 0AA, London, United Kingdom}    
\affil[5]{Leicester Castle Business School, De Montfort University, Gateway House, LE1 9BH, Leicester, United Kingdom}
\begin{document}

\maketitle  
\newpage
\begin{abstract}  
   The so called Benford's laws  are    of  frequent use in order to observe anomalies and regularities in data sets,  in particular, in election results and financial statements. Yet, basic financial  market indices have not been much studied, if studied at all, within such a perspective.

This paper presents features in the   distributions  of S\&P500  daily closing values and the corresponding daily  log-returns over a long time interval,
 [03/01/1950 - 22/08/2014], amounting to 16265 data points.
 We address
the frequencies of the first,  second, and first two  significant digits counts and explore the conformance  to Benford's laws
of these distributions at  five  different (equal size)  levels of disaggregation. 
    The    log-returns are studied for either positive or negative cases.
    The results   for the  S\&P500  daily closing values  are   showing a huge lack of non-conformity, whatever the  different levels of disaggregation. Some ``first digits''  and ``first two digits'' values are even missing.
    The causes of this non-conformity  are  discussed, pointing to the danger in taking  Benford's laws for granted in huge data bases, whence drawing ``definite conclusions''. The agreements with Benford's laws are much   better for the log-returns. Such a disparity in agreements finds an explanation in the data set itself:  the inherent trend in the index.
    To further validate this, daily returns have been simulated
calibrating the simulations with the observed data averages
and tested against Benford's laws.  One finds that not only  the trend but also the standard deviation of the  distributions are relevant parameters in concluding about   conformity with Benford's laws.       
     
\end{abstract}


 \newpage
\section{Introduction}\label{introduction}
\cite{newcomb1881note} noticed that the first few pages of logarithmic table books are more thumbed than the latter ones.  He inferred that numbers with smaller initial digits are more often looked for and used than numbers with larger initial digits. Newcomb   observation  was forgotten for about six decades \citep{mir2018benford}.
 
    In  \cite{benford1938law},  the author apparently and independently\footnote{Benford  does not cite Newcomb. In fact, neither papers   have any bibliography.}   went much ahead in detail and tested the accuracy of his observation by analyzing a large collection of (in total 20000)  numbers,  gathered from twenty diverse fields, thereby  establishing a law as  
 \begin{equation} 
P(d_1)\;=\;log_{10}(\frac{d_1+1}{d_1})=log_{10}(1+\frac{1}{d_1}), 
 \label{eq1BL1}
\end{equation}
  for $ d_1 = 1, 2, 3, \dotsc ,9$, where $P(d_1)$ is the probability of a number having the first non--zero digit $d_1$ and $log_{10}$ is  the logarithm  in base 10. 
  

 The first significant digit of a number is its left-most nonzero digit. According to  Eq. (\ref{eq1BL1}),  
  the  smallest digit, 1,  should  appear as the  first digit with the highest proportion (30.1\%), whereas the  largest digit, 9, will appear as first digit with the least proportion (4.6\%).
 
Thus, 
$N_{d_1}$,  the number of times the integer $d_1 = 1, 2, 3, \dotsc ,9$  is observed to be occurring as the first digit,  in a data set, is given by the so called ``Benford law for the first digit'' (BL1 hereafter) 
\begin{equation}\label{BLeq1}
N_{d_1}= N\; log_{10}(1+\frac{1}{d_1}), \;\;\; d_1 = 1, 2, 3, \dotsc ,9
\end{equation}
where $N$ is the total number of considered   data points. 

One can show that  the probability that $d_2  = 0, 1, 2, 3, \dotsc ,9$ is encountered as  
  the second digit is given by ``Benford law for the second digit'' (BL2 hereafter)
\begin{equation}\label{BLeq2}
 P_2(d_2) =   \sum_{k=1}^{9} \log_{10}\left(1 + \frac{1}{10\;k+d_2}\right) =log_{10}\left[\;\prod_{k=1 }^{9} (\frac{10\;k+d_2+1}{10\;k+d_2})\right]
\end{equation}

Moreover, one can extend BL1 to the first two digits, a so called BL12, 
\begin{equation}\label{BLeq12}
 P_{12}(d_1\;d_2) =  log_{10}(1+\frac{1}{d_1\;d_2}),   \;\;\; d_1\;d_2 = 10, 11, 12,  \dotsc, 98, 99
\end{equation}


 
 Following  a revival due to Nigrini  \citep{Nigrini,NigriniMittermeir}, nowadays,  these so called Benford's laws \citep{bergerhillbook2015,BLreviewMiller2015,mir2018benford} are    of  frequent use in order to observe anomalies and regularities in many data sets \citep[e.g. see][where  widely used survey data sets has been assessed]{kaiser2019benford}. In brief, can one trust the data? 
 
Let us warn that  Benford's laws (BLs) unique origin is not accepted by all theoreticians; in fact,  it might not be unique. Moreover, some discussion rightly exists on whether  Benford's laws should even be valid at all!. One might also discuss how to test the validity (or not) of BLs \cite{DAmico2013, lesperance2016assessing,omerzu2019financial} and \cite{berger2019mathematics} as  examples of such considerations, in the field.
 
 Usually, one considers that  Benford's laws should be valid if there is no data manipulation, or if human constraints are non-existent \citep{hassler2019testing}. Yet, there are cases in which Benford laws are either not found to hold, even though their occurrence should be expected \citep{AusloosHerteliu}, or on the contrary are not expected to be observed, but  are observed \citep{ClippeAusloosTheil2012,TAC2014,CA2015,T2016,ACT2017,shi2018benford}.  Thus, testing BLs on various samples should bring some argument  about discussing the controversies.
 
  Emphasizing the financial statement aspects, it seems strange to us that basic financial  market indices have not been much studied, if studied at all. The section ``State of the Art'' (Sect. \ref{SoA}) allows us to recall what is presently found in the literature concerning the use of Benford's laws for studying financial market  indices. 

 In the present paper, we report our study of Benford's law  first, second, and first two digits validity  (called BL1, BL2, BL12), upon the S\&P500  market closing  values, over a long time interval: from 03/01/1950  till 22/08/2014. This  amounts to 16265 data points. The time series is downloaded from ``Yahoo! Finance'', an authoritative web site providing financial data\footnote{https://finance.yahoo.com}. In doing so we are in line with studies like \cite{Juergens2009,Haley2017,Riccioni2018, shi2018benford}. 
 
 We discuss both daily  closing values and daily log--returns.  Moreover, we divide the whole time interval into five equal size subsets made of 3253 observations each. The interest of such a disaggregation will be explained below.
 
 We observe huge deviations of the  market  closing values  through data histograms with respect to the BLs predictions (or expectations) in Section \ref{dataandanalysis}. The findings are  in disagreement with \cite{Ley1996,Corazzaetal2010,ZhaoWu2010}. We  explain the causes for such a disagreement in Section \ref{conclusions}. Concerning the log--returns, it results that the agreement with BLs is much better; we also explain why. The segmentation of the raw data into 5 time intervals is much serving the explanation. 

 Therefore, even though Benford laws are mainly used to point out to potential frauds in financial statements,  by companies \citep{saville2006using} or countries \citep{rauch2011fact},  one may wonder (or expect) that  such << laws >>  can be used by investors, or in building pertinent models based on volatility. Beside the findings about data ranges, and the role of digits frequency at some  position in the considered numbers,  one may suggest further research, pending  that the considerations can be tied to other techniques based on the frequency of digits, like letters in a text \citep{Shulzinger2017}. This is also relevant to Bayesian approaches (or inputs) and Markov models in investor risk taking aspects.

\section{State of the Art}\label{SoA}



\cite{Ley1996} has apparently been the first to examine ``the peculiar distribution of the US stock indexes' digits''.  
One has to wait 2010, for considerations by Zhao and Wu on the question whether ``Chinese stock indices agree with Benford's Law'' \citep{ZhaoWu2010}.  In both cases, \cite{Ley1996} and \cite{ZhaoWu2010}, Benford's law is claimed to be rather valid. Closely connected to our report,  \cite{Corazzaetal2010} checked whether financial markets  like the S\&P 500 case, from August 14, 1995 to October 17, 2007, thus 3067 data points,obeyed  BL1 \citep{Corazzaetal2010}.   The authors also found some  reasonable agreement, except, they claim,  at anomalous times, like market crashes or special events.
Let us mention \cite{Cinko} and \cite{AybarsAtaunal} where it is tested the ``distribution of BIST-100 returns''  along BLs. More recently,  in 2018, \cite{shi2018benford} looked at whether BL1  could infer the  reliability of financial reports in (6) developing countries. It was shown that ``several visually anomalous data have to be a priori removed'', in order to improve the agreement.

Elsewhere,  i.e. outside market indices studies,  \cite{Rauchetal}  studied  LIBOR Manipulation, performing an ``Empirical Analysis of Financial Market Benchmarks Using Benford's Law''. 
  The authors point to ``a concentration of notably high deviations from the Benford distribution''. 
  In \cite{AlaliRomero}, a decade of financial data  for ``a large sample of U.S. public companies'' is studied along a BL12 perspective. Alali and Romero also  broke ``down the decade of data into six sub-periods'', and found  ``different indicators of manipulation'', similar conclusion against Benford's law compliance are presented in \cite{alali2013characteristics} by the same authors.

In so reading, there is no need to say that more analysis can be welcome, and  subsequent findings  have to be discussed,  as  it follows here below.

  \section{Data and data analysis}\label{dataandanalysis}
  
\begin{figure} 
 \includegraphics[width=\textwidth]
 {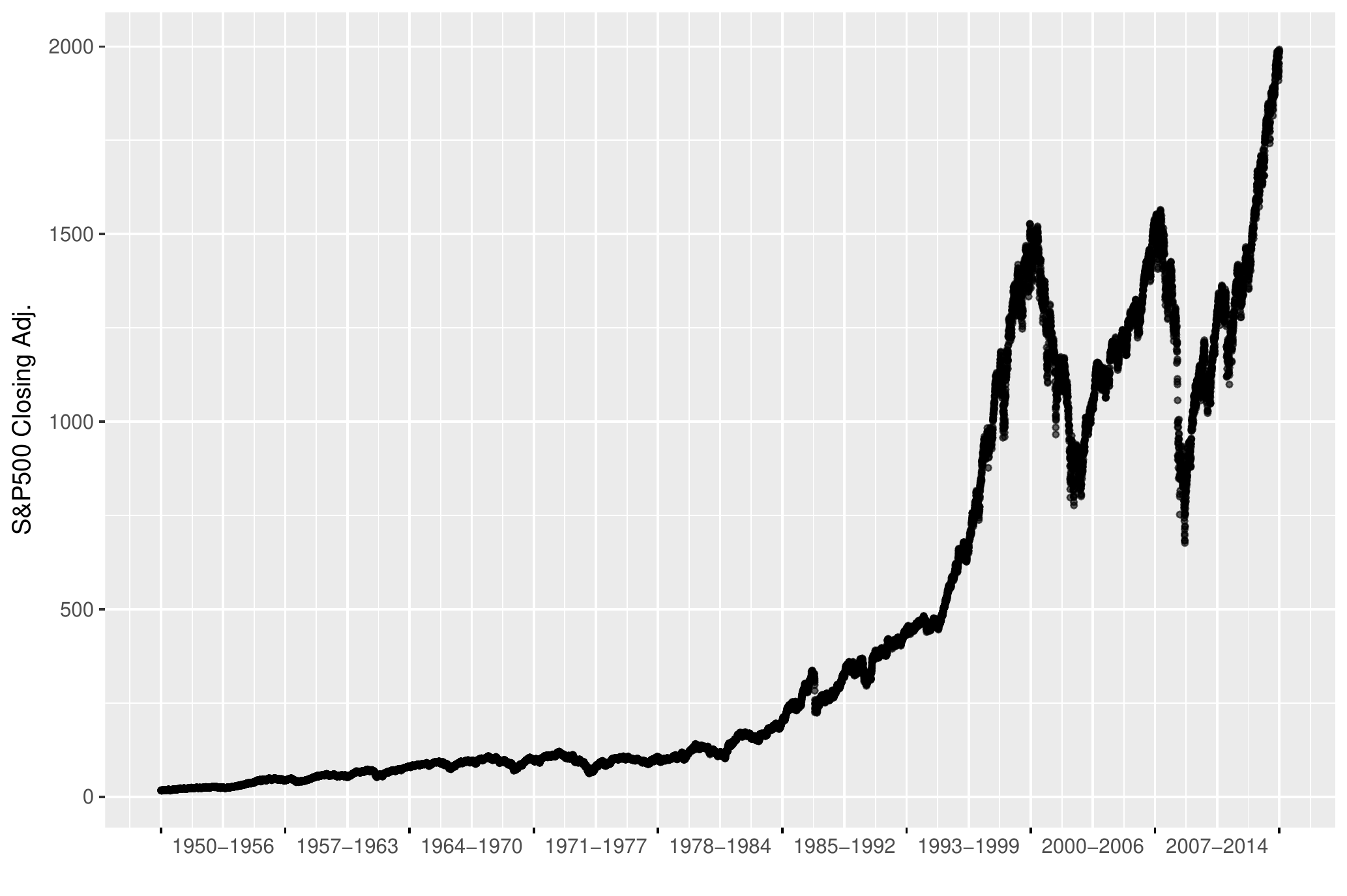}
  \includegraphics [width=\textwidth]
 {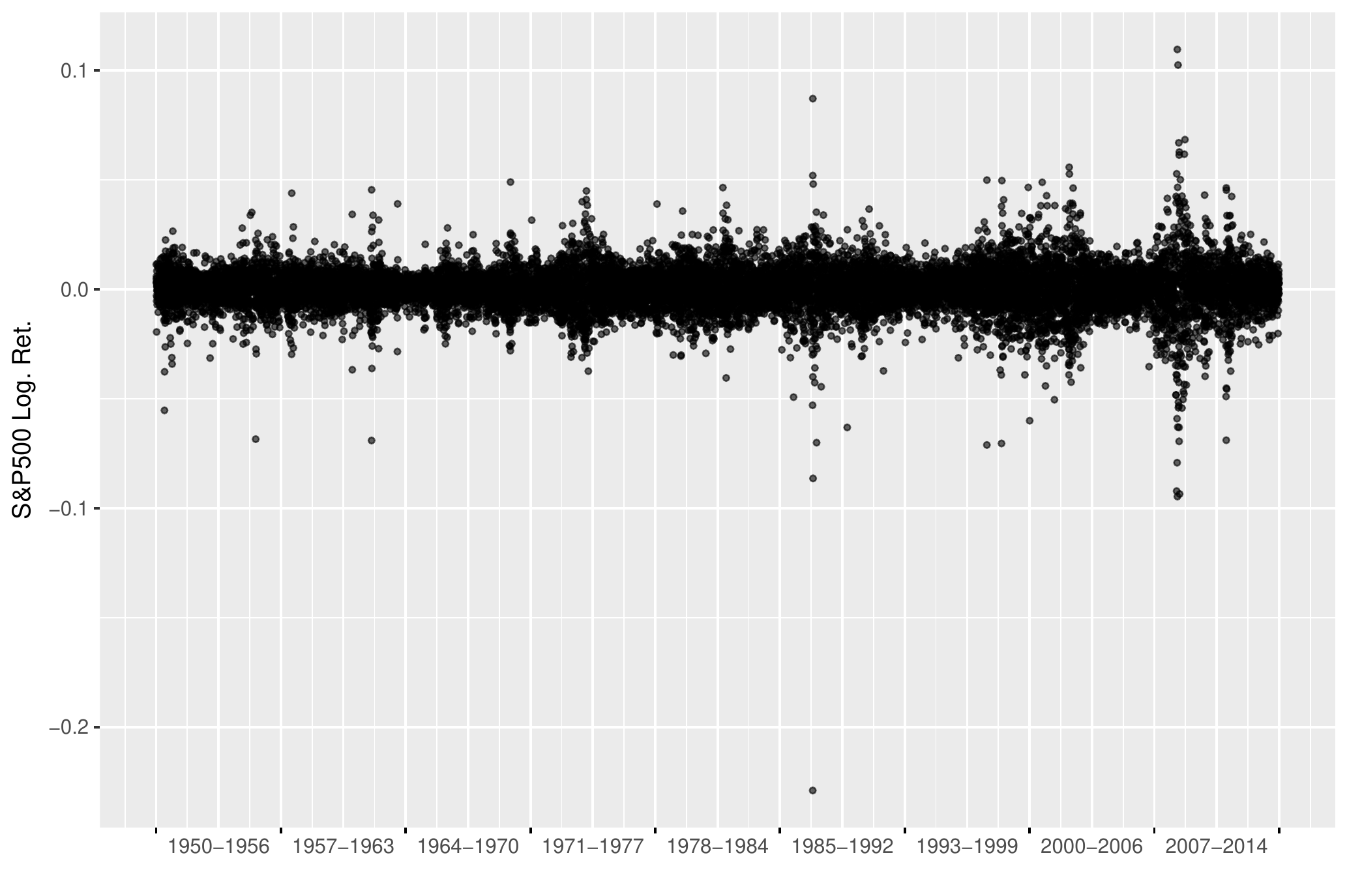}
 \caption[ ] { (top)  S\&P500  daily closing value, (bottom) corresponding  daily log-returns,  between 03/01/1950 and 22/08/2014}
\label{Plot15F1}
 \end{figure}

   We have access to the S\&P500 daily closing  values ($CV$)  via the ``Yahoo! Finance'' web site. The downloaded data cover a period starting on 03/01/1950  and ending on 22/08/2014; it is reported in USD, see Fig. \ref{Plot15F1}. This  amounts to 16265 data points.    From such a set, one can easily obtain the 16264  log-returns ($LR$); see Fig. \ref{Plot15F1} also.  The main  statistical characteristics of such a sample  are reported in Table \ref{Data19T1stats}.   Here it is worth to highlight the huge difference in order of magnitudes  for the S\&P500 closing values, ranging from $\sim 16$ at the beginning of the time series to $\sim 2000$ realized in 2014.

Since there is sometimes some discussions on the adequate  size of the sample \citep{berger2019mathematics} and, for time series, about their ``stationarity'',  we have also divided the original sample into 5 equivalent size groups; thus each set containing 3253 data points.  The corresponding log-returns follow at once.

\begin{table}
\centering 
 \begin{tabular}{|c||c|c|c|c|c|c|c|}

     \hline
 S\&P500  & $CV$& $LR$ \\
\hline
Minimum	&	16.66	&	-0.2290	\\
Maximum	&	1992.4	&	0.10957	\\
N. Points	&	16265	&	16264	\\
Mean	&	451.45	&	 2.9403 10$^{-4}$	\\
Std. Dev.	&	514.08 	&	 9.7315	10$^{-3}$\\
Skewness	&	1.0637	&	-1.0311	\\
Kurtosis	&	-0.32647 	&	27.727	\\
\hline 
\end{tabular}
\caption{ Statistical characteristics of the  S\&P500  (in USD when it applies)  daily closing values ($CV$) and corresponding  log-returns  ($LR$) distributions over the whole data set, [03/01/1950  - 22/08/2014].  The characteristics values are rounded to at most 5 significant digits.
}
\label{Data19T1stats}
\end{table}
  
  A BLs analysis is usually limited to the first, and sometimes second, digit. The second, third and fourth digit distributions are usually found to be rather agreeing with BL2, BL3, and BL4; they  can hardly be used for discussion.   Sometimes, one finds a study of the  first two - BL12 - (and  first three digits, BL123). Thus,  in order to prepare for a BLs analysis, one  usually rounds up the data to at most 5 digits, in order to avoid a rounding of the 4th  significant digit  if it occurs. We kept that rounding rule even though we only consider the first, second, and first two digits, in order to test BL1, BL2 and BL12 on each S\&P500 and log-returns sample. The statistical characteristics of such ``adjusted values'' are presented in Table \ref{DataT2statsallrotated}, for the whole set and for each subsets. The notations seem to be obvious: $CV_k$ and  $LR_k$, with $k= I, ...V$ refer to the subsets. For completeness, let us mention that  the upper limits of such subsets are 3253, 6506, 9759, 13012, and 16265, respectively\footnote{It can be easily understood that we do not take into account the first value of each log-return subset when dividing the whole set into 5 boxes, in order to have the same number of data points, i.e. 3252 for each subset. This is obviously far from a drastic ``assumption''!}.
 
\begin{table}
\centering 
 \begin{tabular}{|c||c|c|c|c|c|c|c|}
  \hline
$ S\&P500$& $Min.$& $Max. $& $Total $&$Mean$ & $ Std.Dev.$&$ Skewness$ &$ Kurtosis$\\\hline
\hline
$CV$	&	16.660	&	1992.4	&	7.34281*	&	451.45	&	514.08	&	1.0637	&	-0.32647	\\ \hline
$CV_I$	&	16.660	&	72.640	&	0.13483*	&	41.447	&	15.828	&	0.076114	&	-1.2629	\\
$CV_{II}$&	62.070	&	120.24	&	0.29463*	&	90.571	&	12.092	&	-0.069025	&	-0.45554	\\
$CV_{III}$&	86.900	&	336.77	&	0.51063*	&	156.97	&	63.706	&	0.96070	&	-0.26424	\\
$CV_{IV}$&	263.82	&	1527.5	&	2.32312*	&	714.15	&	389.79	&	0.69803	&	-1.0539	\\
$CV_V	$&	676.53	&	1992.4	&	4.07961*	&	1254.1	&	258.23	&	0.63624	&	0.30357	\\ \hline \hline
$LR$	&	-0.22900	&	0.10957	&	4.7821	&	 2.9403**	&	 97.315**	&	-1.0311	&	27.727	\\\hline
$LR_I	$&	-0.06909	&	0.04544	&	1.32420	&	4.0721**	&	 74.371**	&	-0.71995	&	8.1493	\\
$LR_{II}$&	-0.03740	&	0.04900	&	0.35396	&	1.0884**	&	 76.573**	&	0.22745	&	3.3794	\\
$LR_{III}$&	-0.22900	&	0.08709	&	1.13920	&	3.5031**	&	 101.43**	&	-3.6266	&	84.952	\\
$LR_{IV}$&	-0.07113	&	0.04989	&	1.26110	&	3.8778**	&	 96.784**	&	-0.35884	&	4.8634	\\
$LR_V$&	-0.09470	&	0.10957	&	0.72213	&	2.2206**	&	 127.70**	&	-0.21277	&	9.1859	\\
\hline 
\end{tabular}
\caption{   Statistical characteristics of  (``top'') the S\&P500 (in USD when it applies) and  (``bottom'') the corresponding daily log-returns 
for the whole data set (16265 and 16264 data points, respectively), and  for the five subsets (3253 and 3252 data points, respectively). N.B. : * $\equiv$  10$^{6}$; ** $\equiv$ 10$^{-4}$. }
\label{DataT2statsallrotated}
\end{table}

One can observe much variety in the data reported in Table \ref{DataT2statsallrotated}:  for example, there are large negative log-returns due to a few crashes, whence  the standard deviation can also be very high.  Also, the skewness and kurtosis, either for the S\&P500 raw data and for the log-returns have different orders of magnitude.

Thereafter, we can compare the number of first, second, and first two digits in such data sets (12, 5 for closing prices, 5 for log-returns and 2 for the global analysis). In the nomenclatures, we distinguish the 5 subsets by different symbols.

Two   warnings first: (i) officially, a zero cannot be a first  digit, when studying BLs; (ii)  decimal points separators are also ignored. Nevertheless, in our counting algorithms, we have kept   0 as a {\it bona fide} first (and also second) digit in the case of BLs tests on  the log-returns.  Indeed, in several (124) cases,   these log-returns are strictly equal to 0, because there was no change in two successive S\&P500  closing values. In such cases, the second digit is of course also 0.
Keeping such a digit, for the tests on log-returns, allows one to observe  the ``relative importance'' of such events; obviously $\leq  \; \sim 1\%$.    
It is easily admitted    that the importance is ``not  great''. However,  ``necessarily'', the number of observed events, $N_k$,  with $k= I, ...V$, thereafter differs in the previously imposed equal size intervals because the zeros are not homogeneously distributed across the 5 log-return subsets.

\begin{figure} 
 \includegraphics[width=0.5\textwidth]  {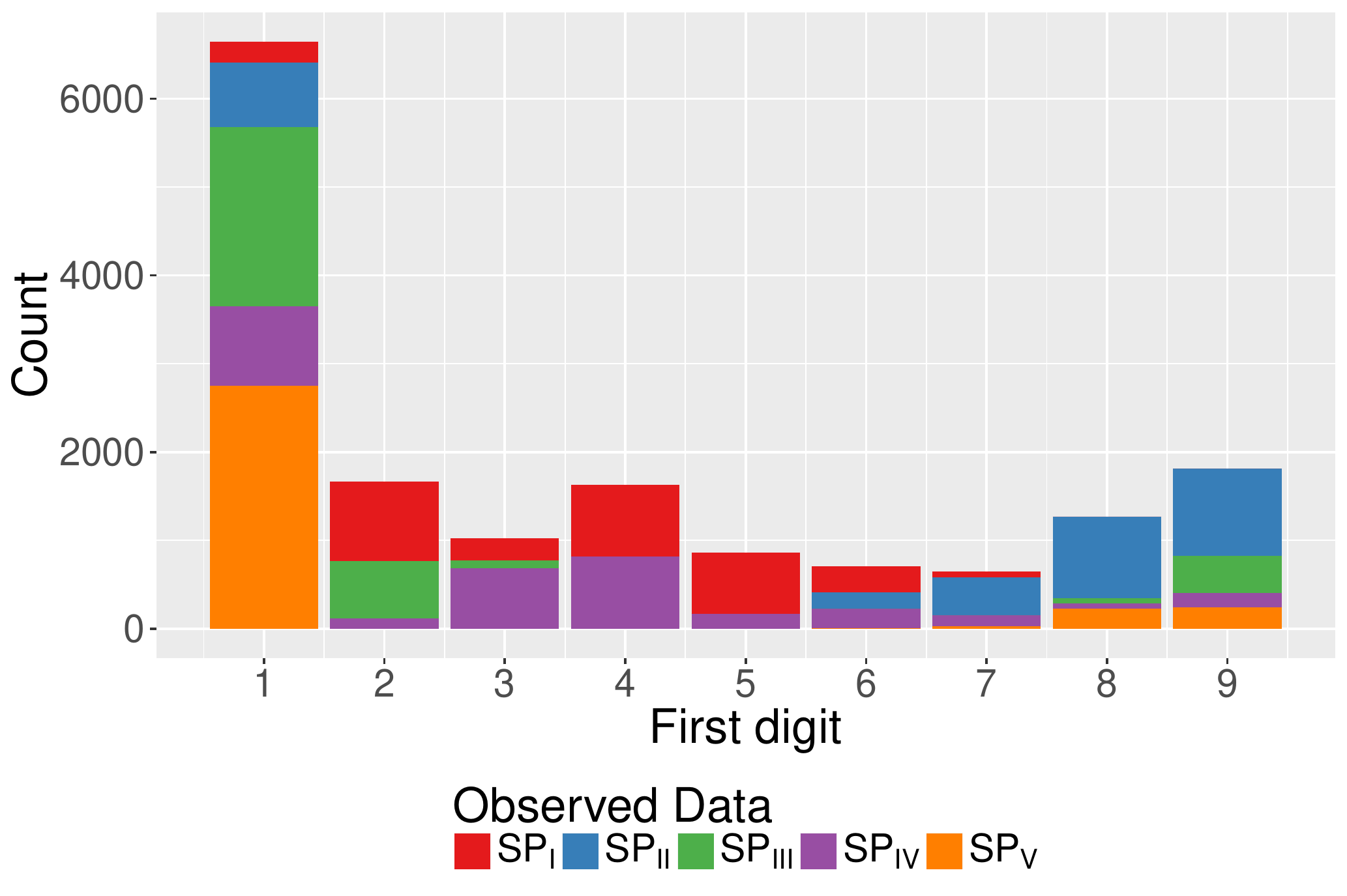}
   \includegraphics [width=0.5\textwidth]  {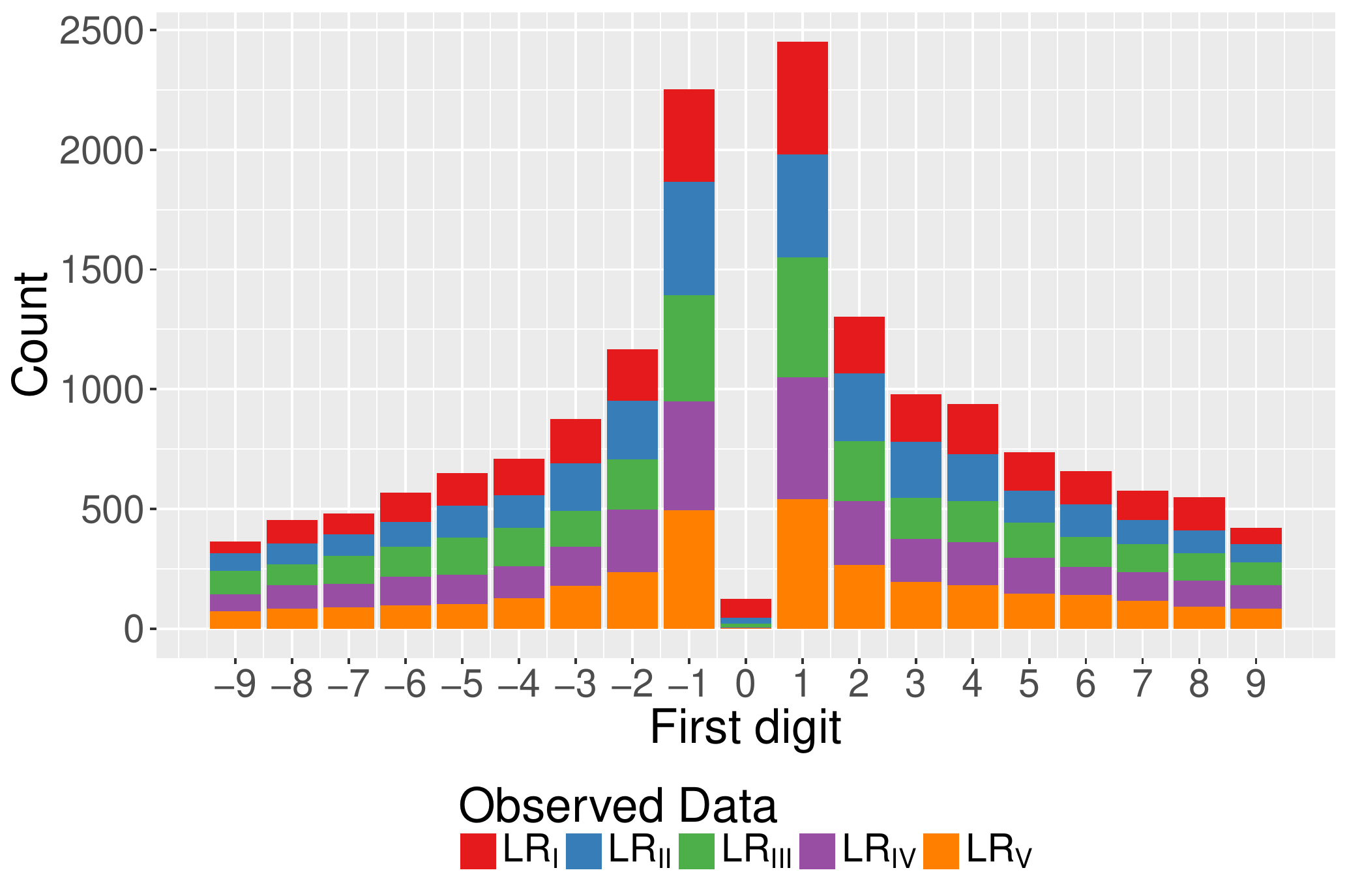}
\caption[ ] {  Test of BL1  on  (left)  S\&P500  closing value, (right) corresponding  daily log-returns  between 03/01/1950 and 22/08/2014}
\label{Plot15F2}
 \end{figure}

Here, we want at once to emphasize the following:  some ``first digits'',  whence ``first two digits'', values are   missing in various subsets. For example, the  missing  first digits in each  $CV_{k}$, can be found  from Table \ref{d1missing}; this is also clearly observed in the  first digit figure for S\&P500, Fig. \ref{Plot15F2}, where one has stacked up the subset histograms.  

 \begin{table}
\centering 
 \begin{tabular}{|c||c|c|c|c|c|c|c|c|c|c|}
  \hline
$d_1$  & 1& 2&3&4&5&6&7&8&9 \\ \hline
 $CV_I$	&	234&898&252&812&689&299&69&0&0 \\
  $CV_{II}$	&  735&0&0&0&0&185&427&921&985	 \\
  $CV_{III}$	&2026&650&93&0&0&0&0&58&426 \\
  $CV_{IV}$	& 899&117&682&818&  171 &220 &126&60&160 \\
  $CV_V$		& 2751&0&0&0&0&4&27&229&242	\\  
\hline 
\end{tabular}
\caption{   Number of $d_1$ digits,   in $CV_{k}$  groups; observe that there are sometimes  missing  digits.}
\label{d1missing}
\end{table}

This is not a trivial point; one understands  (a posteriori) that this is due to the presence of different trends in the data; see the discussion in Sect. \ref{conclusions}
.

\begin{figure} 
 \includegraphics[width=0.5\textwidth]  {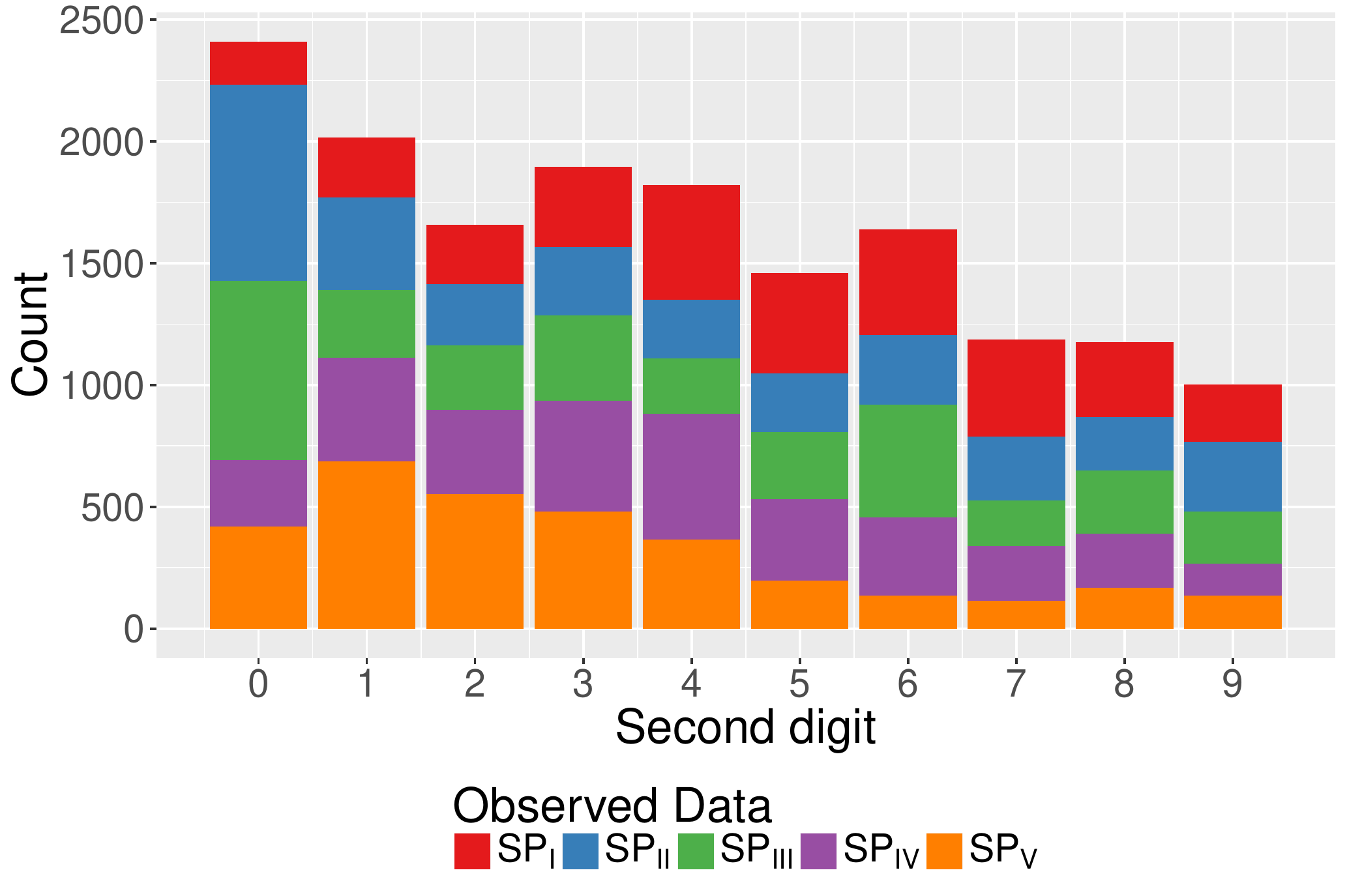}
  \includegraphics [width=0.5\textwidth]  {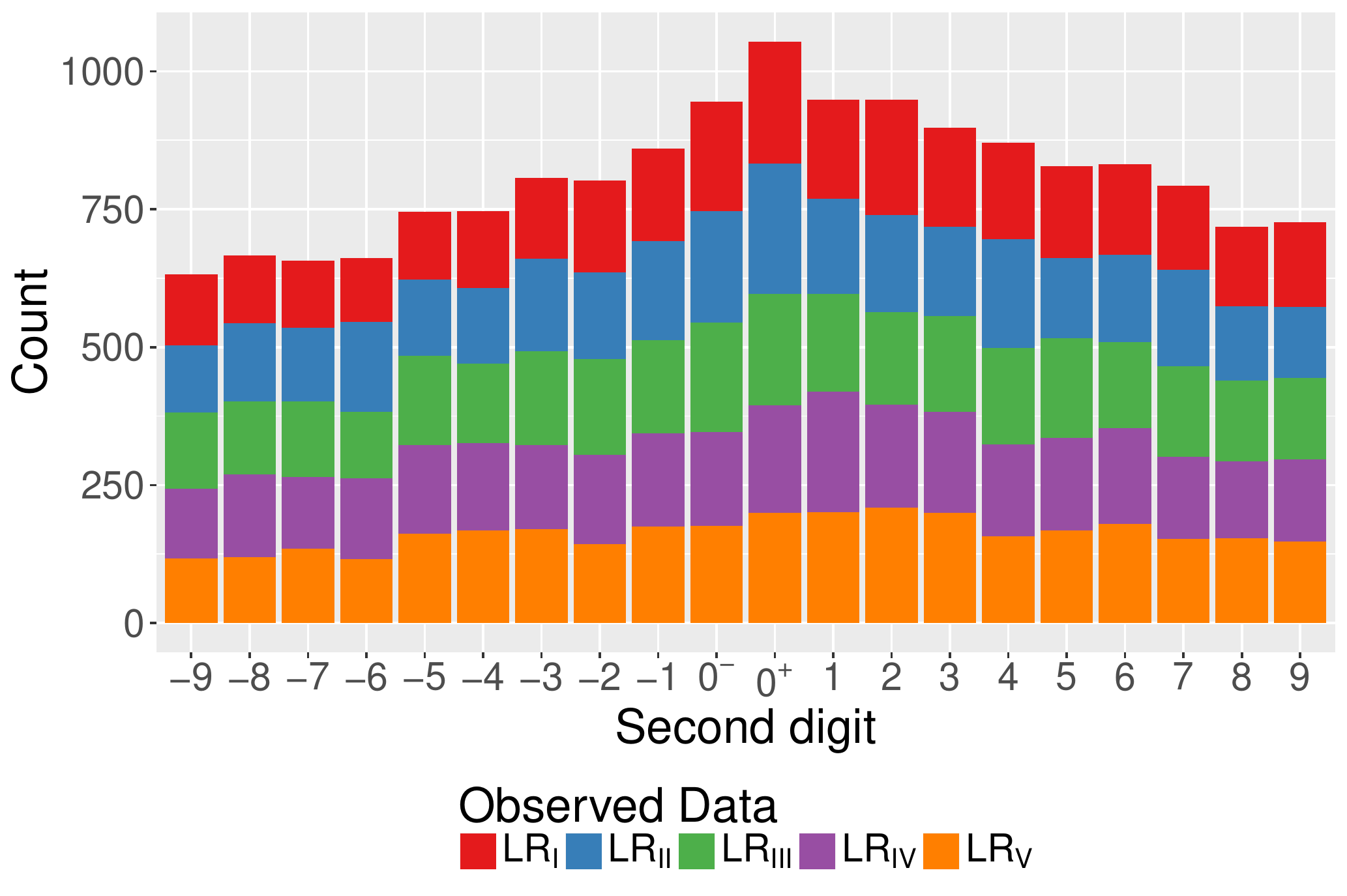}
\caption[ ] {  Test of BL2  on  (left)  S\&P500  closing value, (right) corresponding  daily log-returns  between 03/01/1950 and 22/08/2014}
\label{Plot15F3}
 \end{figure}

\begin{figure} 
 \includegraphics[width=0.5\textwidth]  {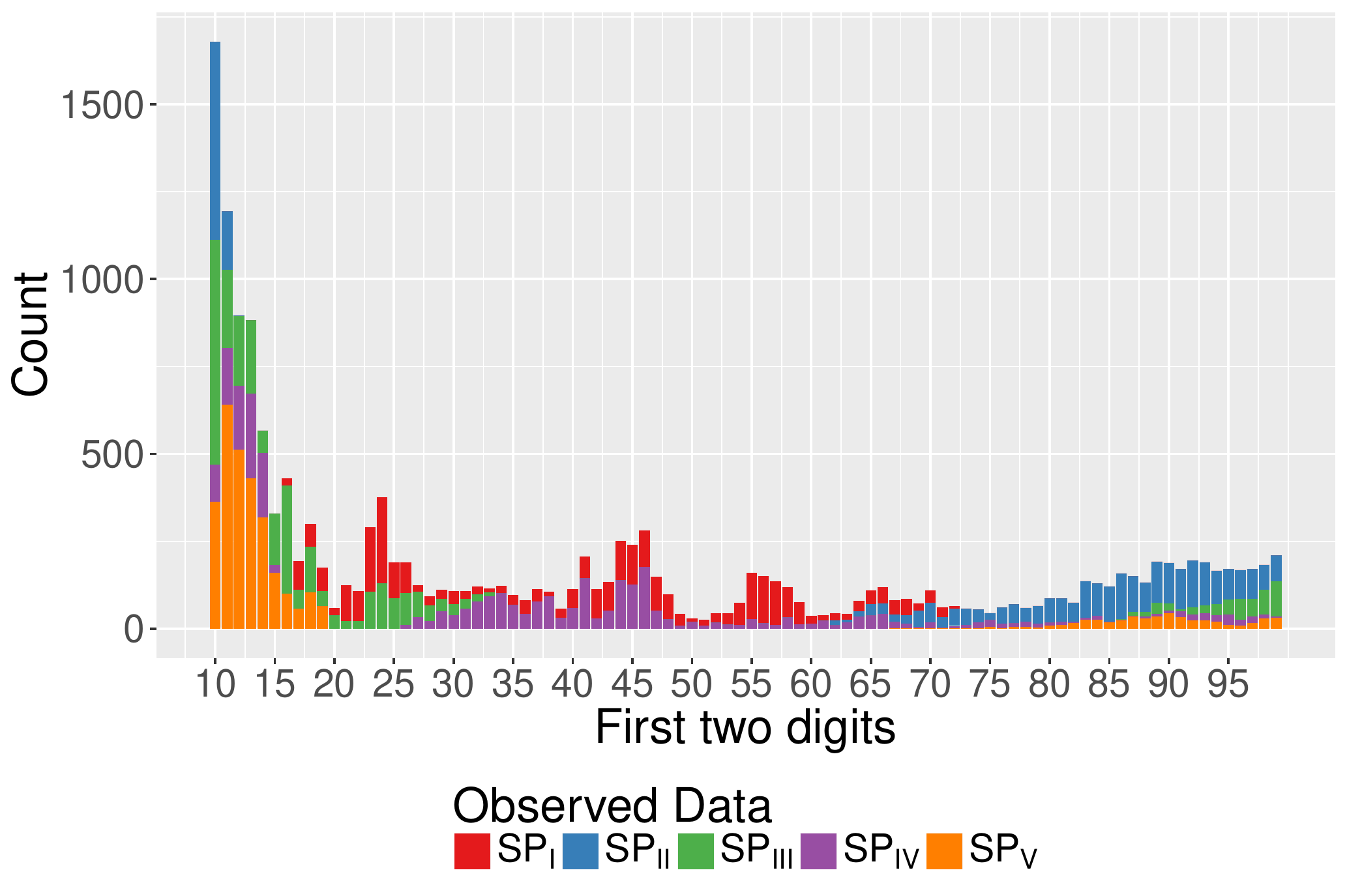}
\includegraphics [width=0.5\textwidth]  {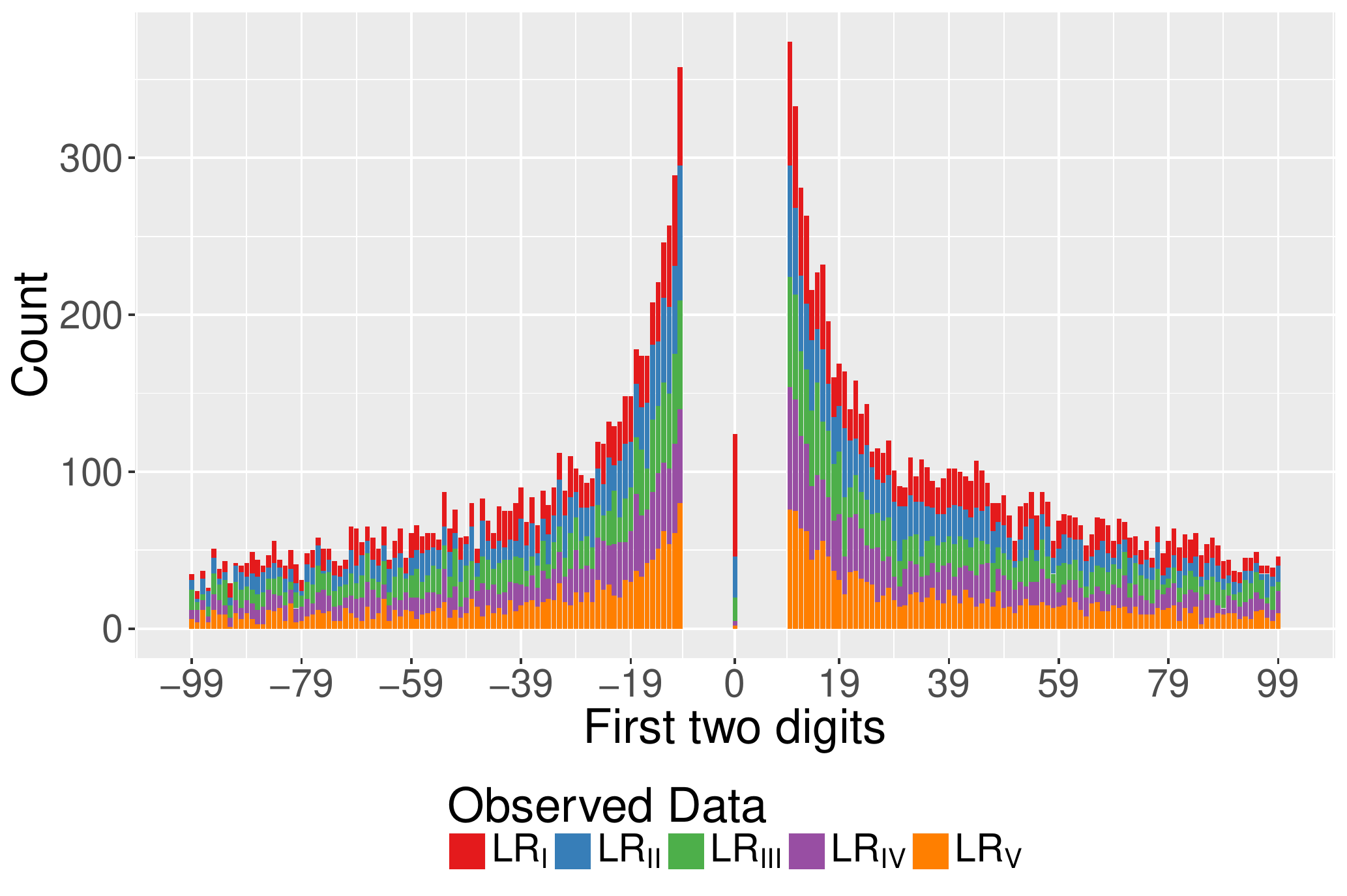}
 \caption[ ] {  Test of BL12  on (left)  S\&P500  closing value, (right) corresponding  daily log-returns  between 03/01/1950 and 22/08/2014}
\label{Plot15F4}
 \end{figure}

  Fig. \ref{Plot15F2} presents the data for testing BL1 on the whole S\&P500 daily closing values and on the corresponding log-returns. The divisions by colour provides information about the examined time intervals. Fig.  \ref{Plot15F3} presents the  corresponding BL2 data, and
   Fig. \ref{Plot15F4}  the data serving for a BL12 analysis.  At once, visually, the S\&P500 data looks hardly 	representable by a log function, like Eqs.(\ref{BLeq1})-(\ref{BLeq12}). In contrast, the log-returns  histograms have a more appealing form. Notice that we distinguish   negative and positive log-returns, and mention on each graph the occurrence of strictly zero and double zero values.

 In Table \ref{T3chi2SP}, we report  the  $\chi^2$ test of variables conformity with BL1, BL2, and BL12 for the S\&P500 whole data set and for the subsets; the number $N_k$ of observations (or data points) is indicated for each case: 16265    and,  for the five subsets, 3253   data points, respectively. The  number of degrees of freedom  ($dof$), easily derived from the number of bins,  is also indicated with the ``critical''   $\chi_c^2 (0.05)$ value.  One can hardly admit any conformity, given the large values compared to the ``critical''   $\chi_c^2 (0.05)$ value.
Even if a  $\chi^2$  test can be claimed as not being the most powerful test for BL conformance \citep{lesperance2016assessing}, the current results are so different from ``any good expectation'' that the utilization of another test will be unlikely able to inverse the conclusions. 
\begin{table}
\centering 
 \begin{tabular}{|c||c||c|c|c|}
  \hline

& &BL1&BL2&BL12 \\ \hline
 &$dof$ :& 8 & 9 & 89   \\ \hline
   $N_k$ & $\chi^2_c(0.05)$:& 15.507 & 16.919 & 112.022    \\ \hline \hline
16265	&$CV$	& 3756.03	& 397.46	&  7084.40 \\ \hline
3253		&$CV_I$		& 2737.22 	& 387.46	&5030.895  \\
3253		&$CV_{II}$	&10038.14	& 544.12	  &12840.86 \\
3253		&$CV_{III}$	& 2936.91	& 527.02	  &5561.862\\
3253		&$CV_{IV}$	& 1476.73	& 302.02 &3496.052\\
3253		&$CV_V$		&5187.90 	& 813.99	 &7664.894	\\ \hline
\end{tabular}
\caption{  Results of  $\chi^2$ test of the daily closing values ($CV$) conformity with BL1, BL2, and BL12 for the S\&P500 whole data set and for the 5 equal size subsets; the number $N_k$ of observations (or data points) is indicated for each case: 16265   for the whole time series and 3253   data points,  for the five subsets,  respectively. The   number of degrees of freedom  ($dof$) is easily derived from the number of bins.  The corresponding   $\chi_c^2 (0.05)$ is given for an immediate comparison. }
\label{T3chi2SP}
\end{table}

\begin{table}
\centering 
 \begin{tabular}{|c||c||c|c|c|}
  \hline
 &   &BL1&BL2&BL12\\ \hline
  &$dof$:& 8 & 9 & 89   \\ \hline
  &$\chi^2_c(5\%)$  &15.507 & 16.919 & 112.022  \\ \hline
 $N_k$ & &  \multicolumn{3}{|c|}{16140   }  \\ \hline
16136  & $LR$ & 156.66 &4.18 & 255.96 \\\hline 	
3174	&$LR_I$& 101.34 & 10.88 & 213.45\\
3226	&$LR_{II}$& 16.61 & 16.30 & 146.46\\
3237	&$LR_{III}$& 86.25 & 8.31 & 172.70\\
3249 	&$LR_{IV}$& 33.99 & 4.49 & 101.39\\
3250	 &$LR_V$	& 19.11 & 5.42 & 102.23 \\
\hline
\hline
  $N_k$ & &  \multicolumn{3}{|c|}{8616}  \\ \hline
8614 & $LR$ & 115.06 & 4.73 & 198.02 \\\hline 
1742	&$LR_{I}^+$ & 74.97 & 5.41 & 168.53 \\ 
1687	&$LR_{II}^+$ & 28.58 & 21.38 & 135.47 \\ 
1690	&$LR_{III}^+$ & 33.30 & 5.87 & 108.90 \\ 
1726 	&$LR_{IV}^+$ & 26.24 & 5.42 & 113.55\\
1769	 &$LR_V^+$	& 17.04 & 7.52 & 91.69  \\
\hline
\hline
  $N_k$ & &  \multicolumn{3}{|c|}{7524}  \\ \hline
7522& $LR$ & 54.46 & 7.01 & 164.81 \\\hline 
1432	&$LR_{I}^-$ & 34.07 & 10.24 & 145.76\\
1539	&$LR_{II}^-$ & 5.59 & 9.53 & 120.75\\
1547	&$LR_{III}^-$ & 59.53 & 8.06 & 154.73  \\
1523 	&$LR_{IV}^-$ & 15.37 & 5.30 & 83.38\\
1481	 &$LR_V^-$	& 12.56 & 13.98 & 123.31 \\
\hline
\end{tabular}
\caption{Results of $\chi^2$ conformity test with BL1, BL2, and BL12 for   the S\&P500   corresponding daily log-returns  ($LR$)
for the    whole data set and for the subsets. The   number of degrees of freedom  ($dof$), easily derived from the number of bins,  is also indicated. The number $N$ of data points is   equal to 16264   for the whole set  and  should be  expected to be 3252  for the five subsets; however,   $N_k$, the ``number of observations'', varies  for  the various cases, since one is not   taking into account the number  (124) of log-return values strictly equal to 0. Moreover, notice that we distinguish (top of table) the case of the absolute values of log-returns and those corresponding to either positive or negative log-return sign (two ``bottom sub-tables''). 
}
\label{T3chi2LR}
\end{table}


Let us turn our attention to the log-returns. As mentioned,  there are 124 cases in which the log-return is equal to 0, since the closing prices are identical two consecutive days; these cases occur unevenly in the different $CV_{k}$ intervals: for completeness, let us mention their occurrence: 78, 26, 15, 3, and 2 times, respectively. This influences the number of observations  $N_k$ in each   $LR_{k}$ subgroup; see first column in Table \ref{T3chi2LR}.

Therefore, there are 16264 - 124 = 16140 cases to be examined in the whole $LR$ series.  When dividing the $LR$ series into 5 subsets, for coherence, the first value in the $II, III, IV,V$, subsets are disregarded, since the first one (day) is ``missing'' in the $LR_I$ case. Thus the number of $LR$ observations  on which to test BL1 amounts to 16260 - 124 =  16136.  

The  number $N$ of data points    should be  expected to be 3252  for the five $LR$ subsets. However,   $N_k$, the number of observations in the $k$-subset, varies  in each subset, since one is not   taking into account the number  (124) of log-return values strictly equal to 0, and such a number is not uniformly distributed through the subsets. Moreover, notice that we distinguish (top of Table \ref{T3chi2LR}) the case of the absolute values of log-returns and those corresponding to either positive or negative log-return sign (two bottom sub-tables). 

The results of the $\chi^2$ test of variables conformity with BL1, BL2, and BL12 for   the S\&P500   corresponding daily log-returns  ($LR$) for the    whole data set and for the subsets are given in Table \ref{T3chi2LR}.   BL1 is hardly obeyed, but the difference between the $\chi^2$  values and the $\chi^2_c$  is not so big as for the $SP$ closing prices sample. Some exceptional cases appear to obey BL1, all of the fall in the study of negative returns, $LR_{II}^-$,  $LR_{IV}^-$ and  $LR_{V}^-$.  The situation is almost perfect for   BL2, for which only  $LR_{II}^+$ is slightly disagreeing. In the case of BL12, only the latest subsets present some agreement, but the first subset and the whole sample series are surely not obeying BL12.

Our explanation follows in the conclusion section. 
\section{A Benford Law compliant price paths generator?}
To stress the dependence from the distributional features of the data against the numerousness of the observations, we test the ability of the standard Geometric Brownian Motion (GBM), \cite{einstein1905molekularkinetischen}, in producing a Benford Law compliant series of returns.  

Along the standard GBM formulation, one has
\begin{equation}
 \ln{\dfrac{S_t}{S_0}} = (\mu - \dfrac{\sigma ^ 2}{2}) * t  + \sigma *  W_t 
\label{GBM}
\end{equation}
where $\ln{\dfrac{S_t}{S_0}}$ is the log-returns, $\mu$ is the mean and $\sigma$ is the standard deviation of the   log-returns, $t$ is the time and $W_t$ is the  Wiener process or Brownian motion. 
Assuming log-normally distributed returns, calibrating the mean via the empirical observations, we aim at finding the level of $\sigma$ that makes the returns simulated via the GBM  as close as possible to the BL1, BL2, BL12 compliance, employing 2 criteria later described.
 
In so doing we have simulated the returns via the following relationship:
\begin{equation}
 \overline{r_{\star,j}} = (\mu_{\star} - \dfrac{\sigma_j ^ 2}{2}) * dt  + \sigma_j *  \sqrt{dt} * \overline{z_{\star,j}}
\label{GBM_sim}
\end{equation}
where $dt$ has been set equal to 1 for convenience without harming the relationship, $\mu_{\star}$ is the average of the returns for the cases $\star = \{LR_{I},LR_{II},LR_{III},LR_{IV},LR_{V}\}$, $\sigma_j$ is the $j-th$ standard deviation from the array ranging from 0.0001 to 0.5 with a step of 0.0001 (the range is set to embed the standard deviations reported in Table \ref{DataT2statsallrotated}). $\overline{z_\star,j}$ is an array made of 5000 random extractions from a $N(0,1)$. Thus, $\overline{r_{\star,j}}$ contains 5000 simulated returns with average $\mu_{\star}$ for each $\star$ and each $j$. Therefore, per each $\star$ of $r_{\star}$ we have a matrix with 5000 rows (simulated days) and 5000 columns (one per each $\sigma_j$). 

\begin{figure}[!tbp]
  \centering
  \subfloat[]{\includegraphics[width=0.48\textwidth]{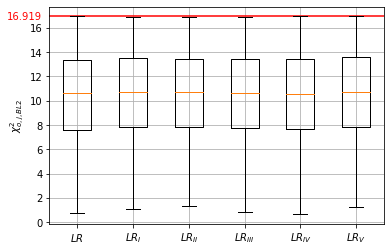}\label{BL2stastsboxplot}}
  \hfill
  \subfloat[]{\includegraphics[width=0.48\textwidth]{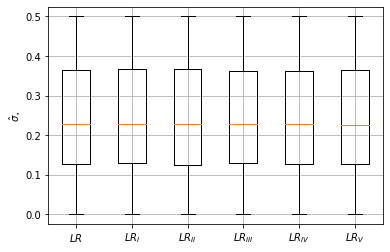}\label{BL2sigmaboxplot}}
  \caption{These graphs contains the results coming form the application of criterion B. (a) is the boxplot of $\chi^2_{o,j,BL2}$ falling below the threshold level 16.919 (red line in the plot) for a $5\%$ significance of the chi square test. So, it represents the  chi square statistics coming from the simulated data and $j$ is the pointer addressing the $j-th$ level of $\sigma_j = [0.0001-0.5]$ used to simulate the data. (b) contains the $\hat{\sigma}_{\star}$ corresponding to the simulated paths having passed the test. Namely, it contains the $\hat{\sigma}_{\star}$ corresponding to the statistics reported in (a)}
\end{figure}

From now on,   $\hat{\sigma}_{\star}$  indicates the standard deviation which produced the most compliant BL price path for the respective $\mu_{\star}$. Therefore, for each column of $r_{\star}$, we calculate the chi-square statistic against the BL theoretical values for BL1, BL2 and BL12 \citep[in doing so we are in line with the usage of the test in comparing disitrubutions, see][]{Christensen1999,Dissanayake2020}. We have determined the target levels $\hat{\sigma}_{\star}$  by using the following 2 criteria separately:   

\begin{enumerate}
\item[A.]   Minimum euclidean distance between the threshold levels of the chi-square distribution at 5\% significance (considering the respective degrees of freedom, see Table \ref{T3chi2LR}) and the observed chi-square levels. The relationship employed is:
\begin{equation}
d_{\star,j} = \sqrt{\sum_{\bullet \in BL}{[\chi^2_{o,j,\bullet} -\chi^2_{c,\bullet}(5\%)]}^2}  \,\,\, \forall	j   \vee  \forall \star 
\end{equation}
Where, $BL = \{BL1,BL2,BL12\}$ representing respectively the stance for first, second and first two digits, $\chi^2_{o,j,\bullet}$ are the observed chi-square statistics and $j$ is the pointer addressing the $j-th$ level of $\sigma_j = [0.0001-0.5]$; $\chi^2_{c,\bullet}(5\%)$ are the threshold taken for the case of $5\%$ significance. For each vector, the $\min(d_{\star,j})$ is reported in Table  \ref{chisigma_1}.

\item[B.] Selection of the $\sigma_j$s for which at least one among BL1, BL2 and BL12 passes the 5\% chi-square test.  The results are reported in Table \ref{chisigma_2}, \ref{sigmastarcrtiB}, Figures \ref{BL2stastsboxplot} and \ref{BL2sigmaboxplot}.
 
\end{enumerate}

For both  criteria used, the results clearly prove that the standard
GBM used for simulating returns makes it impossible to get a joint compliance
with BL1, BL2 and BL12 when  starting from the mean calibrated on real
data. Furthermore, even using 5000 simulated daily returns, one cannot reach
satisfactory results. 

A closer look at the results lead to additional
comments. Under the criterion A (see Table \ref{chisigma_1}), the returns simulated with
the mean of $LR_{III}$ leads to a  $\hat{\sigma}_{LR_{III}}$ = 0.0865 which is very close to the
observed standard deviation for $LR_{III}$, namely 0.0101. However, the $\chi^2$
tests fail for all the digits $d_i$ apart the second, as per the real data  (see Table \ref{T3chi2LR}); in addition, the second digits presents a remarkably low statistics.
The other   $\hat{\sigma}_{\star}$s  are meaningless ; namely they give values rarely met in a financial
Market ;  the $\chi^2$ statistics do not present a relationship
with the sensibleness of the estimations.
   
   The outcomes resulting from the criterion B confirm that the second digits
are the most BL compliant. Table \ref{chisigma_2} shows the number of cases for which
the chi-square statistics pass the test with 5\% significance; it happens in
more than 70\% of the cases for each stance, namely for the majority of the
$\sigma_j \in [0.0001- 0.5]$ plugged in Eq. \eqref{GBM_sim}. Fig. \ref{BL2stastsboxplot} hints about the distributions of
the statistics whose frequencies are reported in Table \ref{chisigma_2}. The $\hat{\sigma}_{\star}$ obtained when   applying the criterion B are summarized in Table \ref{sigmastarcrtiB} and showed in Fig. \ref{BL2sigmaboxplot}.
Most of them are pretty high as testified by the mean and the standard
deviation reported in the summary statistics and in the box.

Summarizing, the results show sensitivity to $\sigma$ and to the presence of trends in the data. Besides, the $d_i$  behaviors are different, therefore, per each digit studied against the respective Benford's law, dedicated consideration should be run before grasping conclusions on the data.

\begin{table}
\centering 
 \begin{tabular}{c|c|c|c|c|c|}
  \hline
 \multicolumn{3}{r|}{ }   & BL1&BL2&BL12\\ \hline
 \multicolumn{3}{r|}{ $dof$:}&  8 &9 &89   \\ \hline
  \multicolumn{3}{r|}{ $ \chi^2_c(5\%)$  }&15.507 & 16.919 & 112.022  \\ \hline \hline
  $\star$ & $\hat{\sigma_\star}$ & eucl. dist. & BL1&BL2&BL12 \\ \hline
 $LR$ & 0.2561 & 210.30 &159.29 & 12.17 & 265.42\\ \hline
 $LR_I$ & 0.2428 & 205.41 &163.80 & 5.39 & 253.69 \\ \hline
$LR_{II}$ & 0.2318 & 201.60 &157.44 & 10.01 & 255.02 \\ \hline
$LR_{III}$ & 0.0865 & 204.60 &127.21 & 6.89 & 283.15 \\ \hline
$LR_{IV}$& 0.2081 & 195.18 &152.88 & 13.01 & 250.63 \\ \hline
$LR_{IV}$ & 0.2382 & 193.53 &158.17 & 7.72 & 242.47 \\ \hline
\end{tabular}
\caption{Results of $\chi^2$ test of variables conformity with BL1, BL2, and BL12 for returns simulated according  to Eq.\eqref{GBM_sim} and with $\hat{\sigma_\star}$, which is the standard deviation estimated to satisfy the criterion A for each row. The   number of degrees of freedom  ($dof$), easily derived from the number of bins.}
\label{chisigma_1}
\end{table}

\begin{table}[htbp]
  \centering
    \begin{tabular}{c|c|c|c|}
  \hline
   & BL1&BL2&BL12\\ \hline
  $dof$: &  8 &9 &89   \\ \hline
  $ \chi^2_c(5\%)$ &15.507 & 16.919 & 112.022  \\ \hline \hline
    $LR$  & 0     & 3596  & 0 \\
    \hline
    $LR_I$ & 0     & 3610  & 0 \\
    \hline
    $LR_{II}$ & 0     & 3639  & 0 \\
    \hline
    $LR_{III}$ & 0     & 3619  & 0 \\
    \hline
    $LR_{IV}$ & 0     & 3640  & 0 \\
    \hline
    $LR_{V}$ & 0     & 3576  & 0 \\
    \hline
    \end{tabular}%
\caption{Number of times that $\chi^2$ test for verifying conformity with BL1, BL2, and BL12 for the simulated returns have returned acceptable results. Namely, with the returns simulated thanks to the relationship \eqref{GBM_sim} and with $\sigma_\star \in [0.0001 - 0.5]$, we can see the number of times that the criterion B have been respected for each stance (row). Hence, the figures represent the frequencies of the chi-square statistics being below the threshold  $ \chi^2_c(5\%)$. }
\label{chisigma_2}
\end{table}

\begin{table}[htbp]
  \centering
    \begin{tabular}{c|c|c|c|}
    \hline
          & $\mu$ & $\sigma$ & $\mu / \sigma$ \\
    \hline
    $LR$  & 0.2417 & 0.1441 & 1.6778 \\
    \hline
    $LR_I$ & 0.2423 & 0.1435 & 1.6886 \\
    \hline
    $LR_{II}$ & 0.2419 & 0.1450 & 1.6688 \\
    \hline
    $LR_{III}$ & 0.2427 & 0.1431 & 1.6967 \\
    \hline
    $LR_{IV}$ & 0.2418 & 0.1436 & 1.6837 \\
    \hline
    $LR_{V}$ & 0.2420 & 0.1442 & 1.6786 \\
    \hline
    \end{tabular}%
  \caption{This is a statistical summary of the $\hat{\sigma}_{\star}$ resulting from the application of the criterion B. Namely, we report the mean, the standard deviation and the variation coefficient of the resulting $\hat{\sigma}_{\star}$.  From Table \ref{chisigma_2} one can see that just the BL2 has passed cases, therefore, the statistics here reported concerns the $\hat{\sigma}_{\star}$ resulting from the passed tests for the second digit of the simulated data. The boxplots in Fig. \ref{BL2sigmaboxplot} report additional information about $\hat{\sigma}_{\star}$ for each considered stance.}
  \label{sigmastarcrtiB}%
\end{table}%

\section{Conclusions}\label{conclusions}

In view of increasing knowledge about  applications of Benford's laws, we have analyzed 
  features in the   distributions  of S\&P500  daily closing values and the corresponding daily  log-returns over a long time interval, that is,  from
the first days of January 1950  till almost the end of August 2014, amounting to 16265 data points.
We have addressed our considerations to the  amount of first,  second and first two  significant digits. We have also explored the conformance  to Benford's laws
of these distributions  distinguishing five  different (equal size)  levels of disaggregation, in order to test some (non)stationarity (hidden)  feature, - if it might occur. Moreover, although this is not usual, we have distinguished  negative log-returns from positive ones, plus their combination, since we have enough available data points.
   
 The results   for the  S\&P500  daily closing values  ($CV$) are unexpectedly  showing a huge lack of conformity, whatever the  different levels of disaggregation. We have noticed that some ``first digits''  and ``first two digits'' values are  missing in some subsets.  
The agreements with Benford's laws are much   better for the log-returns ($LR$). Such a disparity in agreements finds an explanation in the data set itself, rather than in a possibility of fraud!

In fact, this feature allows us to comment on some often forgotten criterion for  testing the conformity of BLs \citep{berger2019mathematics}. Indeed, one should emphasize that BLs could only be usefully studied and observed  if all digits - from 1 to 9 - are well represented as every first digit. A time series or a set of data points should first be tested for its range, basically, the minimum and maximum values.  The argument is here well sustained by observing the evolution of the S\&P500  $CV$  over time. 

\begin{figure} 
 \includegraphics[width=\textwidth]
 {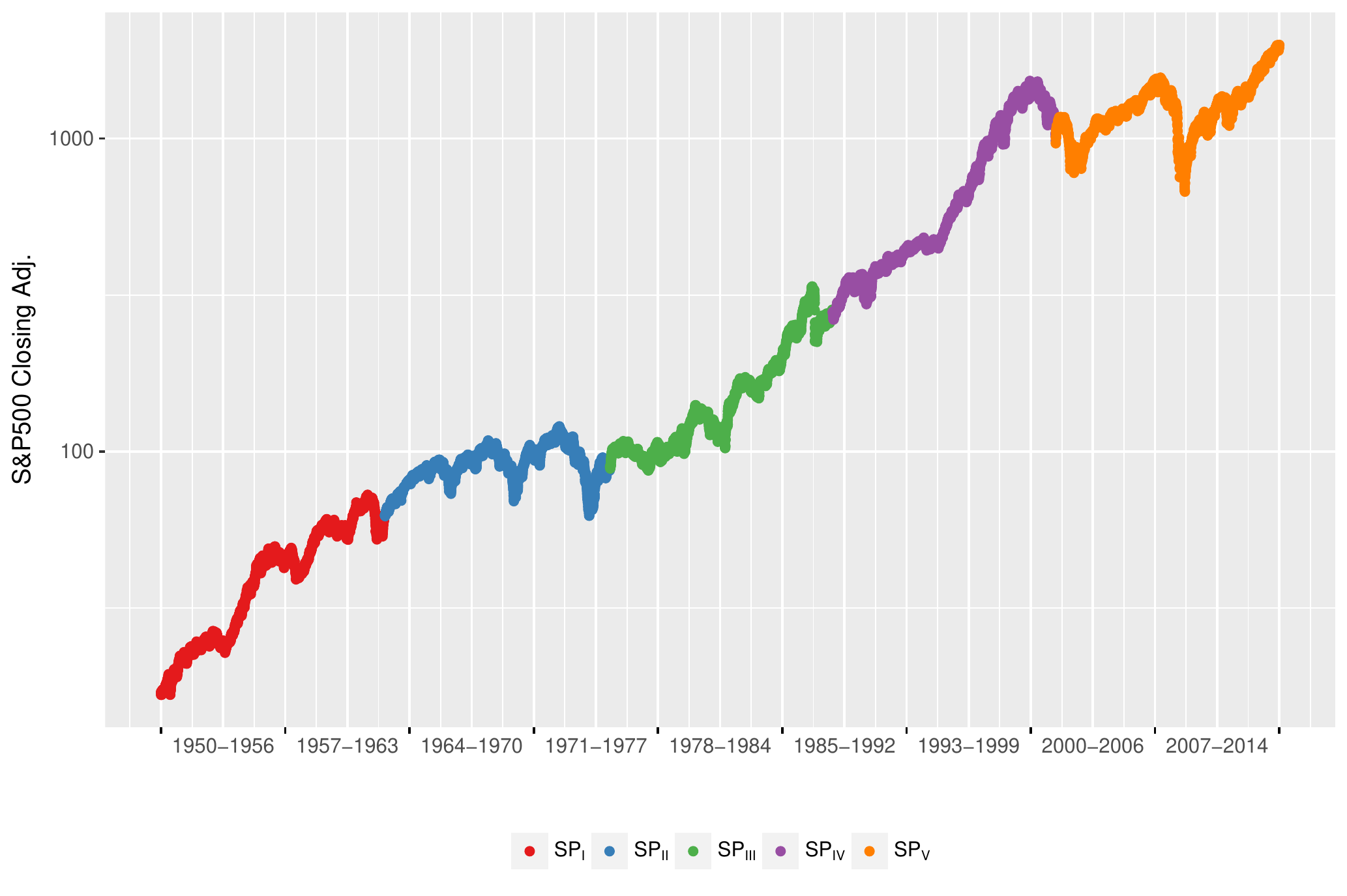}
\caption[ ] {  Semilog plot of S\&P500  closing values  between 03/01/1950 and 22/08/2014 emphasizing  studied subsets}
\label{Plot2logSPsectors}
 \end{figure}

Fig. \ref{Plot2logSPsectors}  provides a semi-log view of the S\&P500 closing values;  the five studied subsets are emphasized. This allows one to understand why the distributions of digits are peculiar. Having ``abandoned'' a 1 first digit in some sector due to the financial trend, it takes ``a while'' before one goes from a 9 to a new 1 (for the following decade for example).  Another  example showing why sometimes a BL analysis and  anomaly deduction  might be doubtful is found in sector $CV_{II}$: the index starts from $\simeq$ 62, reaches $\simeq$ 89, but never goes to any 50, or 20 or 200,  a fortiori  300, etc.  Thus, the index ``misses'' a few first digit values. The same observation goes true for the other sectors where first digits are missing. In the present analysis of a financial market, this is due to the inherent trend in the index. Such causes for no conformity explains previously puzzling observations like in \cite{ClippeAusloosTheil2012}. Related explanations do  follow for cases of data containing crashes, and ``long time'' spent in growing and recovering \citep{Corazzaetal2010}. 
   
    Thus, beside a thorough analysis of a financial index, a case rarely examined, surely over a so big set of data points,  specific causes of this non-conformity  are  presented, pointing to the danger of taking  Benford's laws for granted in huge data bases, whence leading to ``definite conclusions''.    

One often reads ``the more, the better'' as in \cite{Narin2011} or \cite{Collins2017} where it is claimed ``the larger, the better''  for applying Benford's laws and deducing fraud or not through  lack of conformity or not. This is not true! A large set of data points is  neither a sufficient nor necessary criterion for such a statistical conformity test  \citep{heilig2018testing}. Under this perspective, we have simulated
5000 daily returns using the averages of the real data presented
in Section 
\ref{dataandanalysis} with the Geometric Brownian Motion formulation, see 
Eq.\eqref{GBM_sim}.
 In addition, for each mean, hence for each studied time span, the
standard deviations plugged in  Eq.\eqref{GBM_sim} range in [0.0001 - 0.5]
(with steps of  0.0001). The BL compliance results are in line with the
outcomes obtained with the observed data. This is an additional
hint; in fact, one needs to consider the distributional features of
the phenomenon under investigation instead of only  focusing on the
number of observations  or on the granularity of the data.
This type of comments is  in line with \cite{mebane2011comment}, where the
author has commented comparable exercise runs for studying fraud
detection application of BL in political elections.

 Finally, recall that BLs are used to detect fraud mainly. Of course, there are data sets which can be hardly manipulated. We are all convinced that S\&P 500 and other  financial indices result from averages, thus apparently  obeying the BL validity theoretical criteria, whence could not have fraudulent aspects. However, the present study suggests that one might use BLs at a more << microscopic level >>, that of company share price, as already appreciated by \cite{saville2006using}.

As already stated, one of the main findings of our research has been about the data range. Indeed to conform with the law, the data set must contain data in which each number 1 through 9 has an equal chance of being the leading digit; there should be equipartition \citep{janvresse2004uniform, iafrate2015equipartitions}.  However, this seems paradoxical. What we show is that the data transformation, from the raw index value to the log-return space,  is a key step for observing that there is no data manipulation and obedience to BL. The trend value is avoided. Moreover,  BL2 and BL12  are less sensitive to trend manipulation.

As so observed,  one may imagine that BL2 and BL12 are of interest for investors, since a change of the first digit is rather rare  when share prices are higher than 10 (whatever the currency is in fact).  BL1 should be verified for << cheap prices >>, lower than 10. This would lead to an investment strategy similar to that considering  the equivalence of digits in data series to letters in texts \citep{ausloos2003strategy}. Whence it would be interesting for financial analysts to reconsider a connection between  Benford and Zipf law approaches. 

\clearpage
 
\clearpage


\end{document}